\documentclass[%
 amsmath,amssymb,
 aps,
prf,
]{revtex4-2}

\usepackage{graphicx}% Include figure files
\usepackage{dcolumn}% Align table columns on decimal point
\usepackage{bm}% bold math
\usepackage{color,graphicx}% color
\usepackage[dvipsnames]{xcolor}
\usepackage{ulem}
\usepackage{epstopdf, epsfig}
\usepackage{subcaption}
\usepackage{dashrule}
\usepackage{multirow}
\usepackage{natbib,url,hyperref}
\usepackage{amsmath}
\usepackage{lipsum}% for dummy text
\usepackage{tabularx}
\usepackage{rotating}
\usepackage{hyperref}% add hypertext capabilities
\usepackage{color}
\usepackage{xcolor}

\definecolor{grey}{rgb}{0.7,0.7,0.7}

\definecolor{cNeutralGray}{RGB}{99,101,105}
\definecolor{cGreenCurrent}{rgb}{0.4660,0.6740,0.1880}
\definecolor{cEverGreen}{rgb}{0,0.5,0}
\definecolor{cCyanCurrent}{rgb}{0.3010, 0.7450, 0.9330}
\definecolor{cBlueSecondary}{RGB}{0, 75, 135}
\definecolor{cViolet}{RGB}{204, 0, 204}
\definecolor{cYellow}{RGB}{255, 255, 51}
\definecolor{cOrangeUtility}{RGB}{242, 160, 0}
\definecolor{cRedUtility}{RGB}{183, 49, 44}

\definecolor{cBlueBrand}{RGB}{47, 126, 178}
\definecolor{cVioletCurrent}{rgb}{0.4940, 0.1840, 0.5560}
\definecolor{cBlack}{rgb}{0, 0, 0}
\definecolor{cYellowCurrent}{rgb}{0.9290, 0.6940, 0.1250}

\begin{document}

% \preprint{APS/123-QED}

\title {Reynolds number required to accurately discriminate between\\
trends of skin friction and streamwise peak intensities in wall turbulence}

\author{Hassan Nagib}%
\email{nagib@iit.edu}
\affiliation{%
ILLINOIS TECH (IIT), Chicago, IL 60616, USA
}%

\author{Peter Monkewitz}%
\email{peter.monkewitz@epfl.ch}
\affiliation{%
\'Ecole Polytechnique F\'ed\'erale de Lausanne (EPFL), CH-1015 Lausanne, Switzerland
}%

\author{Katepalli R. Sreenivasan}%
\email{katepalli.sreenivasan@nyu.edu}
\affiliation{%
Tandon School of Engineering, Courant Institute of Mathematical Sciences, Department of
Physics, New York University, New York, USA
}%

\date{\today}% It is always \today, today,

\begin{abstract}
We expand on the conclusion of Nagib, Chauhan \& Monkewitz~\cite{NCM07} that nearly all available skin friction relations for zero-pressure-gradient turbulent boundary layers come into remarkable agreement over the entire range $Re_\theta$ $<$ O($10^8$), where $Re_\theta$ is the momentum thickness Reynolds number, provided one coefficient is adjusted in each relation by anchoring it to the most accurate measurements known. Regarding the peak of the streamwise turbulence intensity %$<uu>^+_P$, 
we find good agreement between the three analyses of Monkewitz \cite{M22}, Chen \& Sreenivasan \cite{CS22} and Monkewitz \& Nagib \cite{MN15}, with some coefficients slightly modified. We conclude that, several existing comparisons notwithstanding, accurate measurements in flows with $Re_\tau$ $ > $O($10^6$), where  $Re_\tau$ is the friction Reynolds number, are required to {\it directly} discriminate between the existing theories of Marusic \& Monty (2019) \cite{MM19} and Chen \& Sreenivasan \cite{CS22}. 
\end{abstract}

% TEXT FOR TEASER (A TEASER WILL BE NEEDED, and a figure)
% 
% FIGURE FOR TEASER IS wing2bis.png
\maketitle
%\tableofcontents
Purpose of the note: Our goal here is to add to the discussion on the important topic of the asymptotic behavior of streamwise turbulent stress peaks in the wall layer. In particular, we estimate the Reynolds numbers needed for a \textit{direct} assessment of the two competing theoretical results for the behavior of the variance. To do this properly, we first need to ascertain the accuracy of available measurements of skin friction and the Reynolds number. Accordingly, we organize the paper into two sections---accuracy of available skin friction data, and the mean square values of streamwise fluctuations in the wall layer---and present a conclusion.

\begin{figure}
      \centering
      \includegraphics[width=0.7\textwidth]{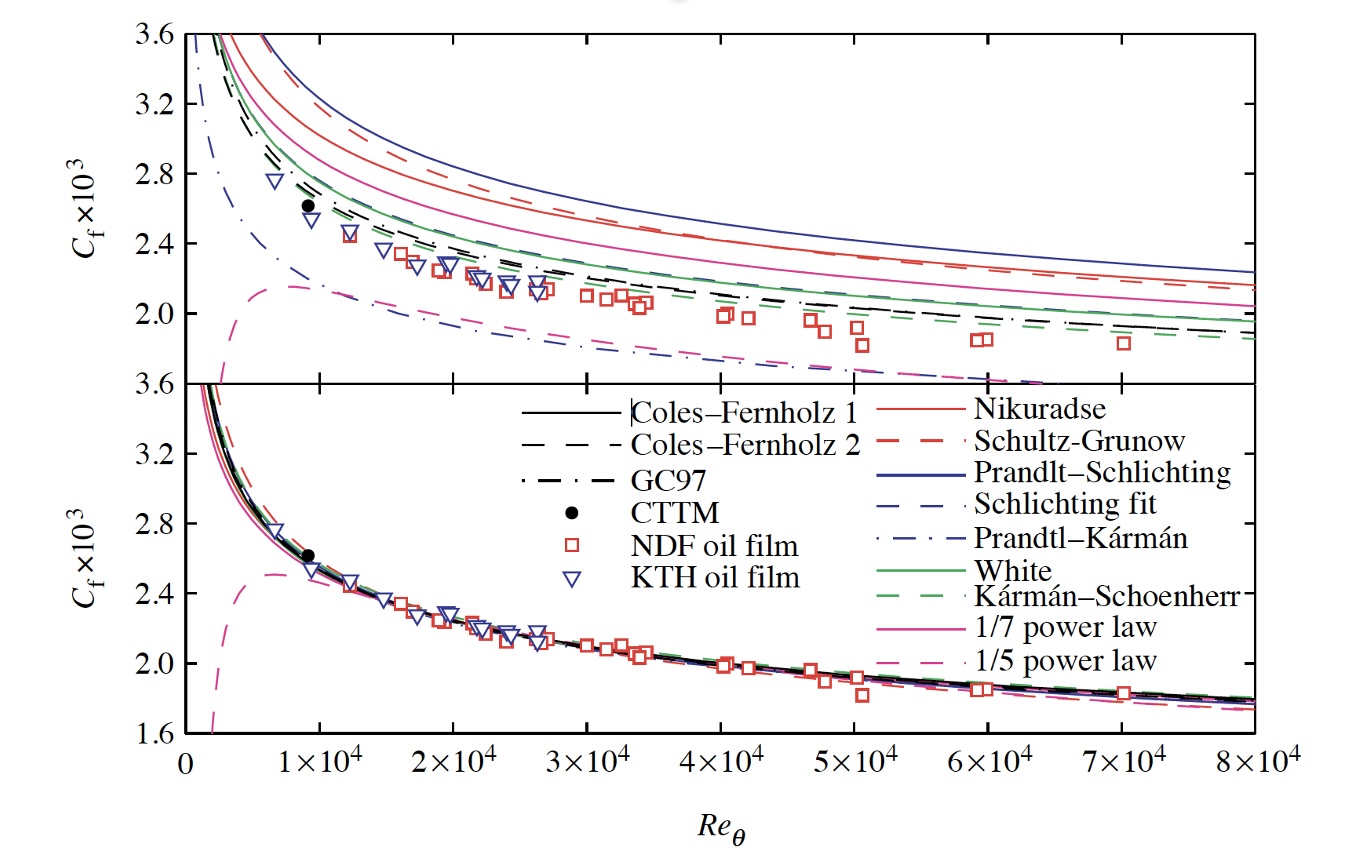}
       \caption{Reproduction of figure 1 of Nagib, Chauhan \& Monkewitz (2007) \cite{NCM07}: Variation of skin-friction coefficient with Re$_\theta$. Experimental data from NDF and KTH are compared with relations for ZPG TBLs found in the literature.}
       \label{fig:fig1}
\end{figure}

Skin friction in turbulent boundary layers: For over a century, several relations have been proposed to describe the skin friction coefficient in zero pressure gradient (ZPG) turbulent boundary layers (TBLs).  The differences between them can be as much as 20\% to 40\% as reflected in the top part of figure \ref{fig:fig1}.  In most cases these empirical correlations are not based on rigorous analysis and often lack even the empirical underpinning by accurate experimental data, or are based on observations from boundary layers that are not in strict ZPG conditions. About two decades ago the availability of the oil film interferometry to measure wall-shear stress with an accuracy in the order of 1.5\% has allowed measurements in the facilities at ILLINOIS TECH (IIT), Chicago, and in KTH, Stockholm, over wide range of Reynolds numbers in accurately established ZPG conditions---i.e., with variations of freestream velocity of the order of 0.2\%. The data from these two experiments are included in the present figure \ref{fig:fig1} and in Nagib, Chauhan \& Monkewitz (2007) \cite{NCM07}, and none of the existing relations in the literature is in good agreement with the data.

\begin{figure}
      \centering
      \includegraphics[width=0.7\textwidth]{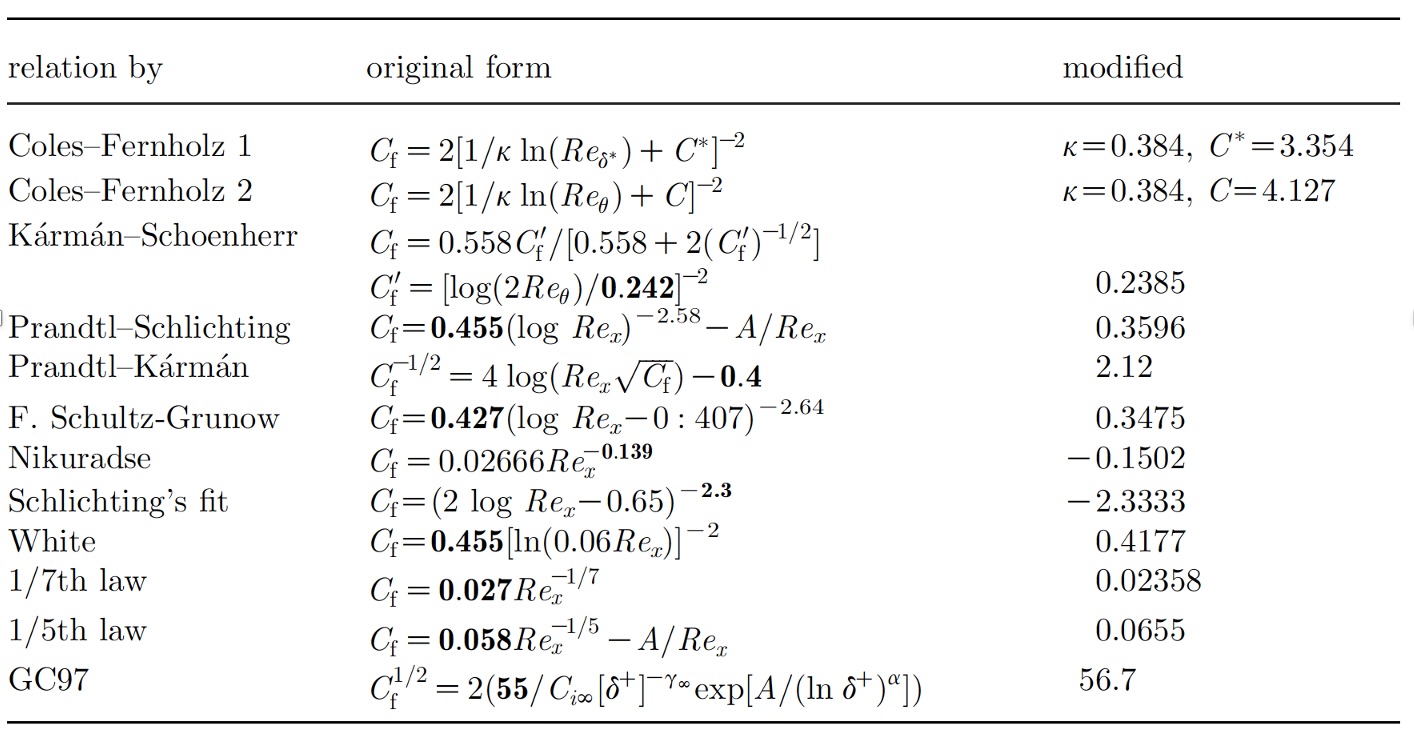}
        \caption{The figure reproduces an image of table 1 of Nagib, Chauhan \& Monkewitz (2007) \cite{NCM07}: Skin-friction relations used to fit the KTH and NDF oil-film data in figure 1. (Throughout the table, $x$ is the downstream distance measured from the leading edge.)}
        \label{fig:fig2}
\end{figure}

A table reproduced as figure \ref{fig:fig2} summarizes the various skin friction coefficients $C_f$, found in the literature as a function of the momentum thickness Reynolds number, $Re_{\theta}$, identified as ``original form''.  Fitting each of the relations to the measured skin friction using oil film interferometry, while allowing only one coefficient, marked with bold text in the figure, adjusted using a least-squares algorithm, yields the ``modified'' coefficients listed in the table.  Note that the Coles-Fernholz relations were fitted directly to the data, with the coefficient $\kappa$ and the additive constant included in the figure. The bottom part of figure \ref{fig:fig1} shows remarkable agreement between all the ``modified'' relations including the Coles-Fernholz relations. 

\begin{figure}
      \centering
      \includegraphics[width=0.7\textwidth]{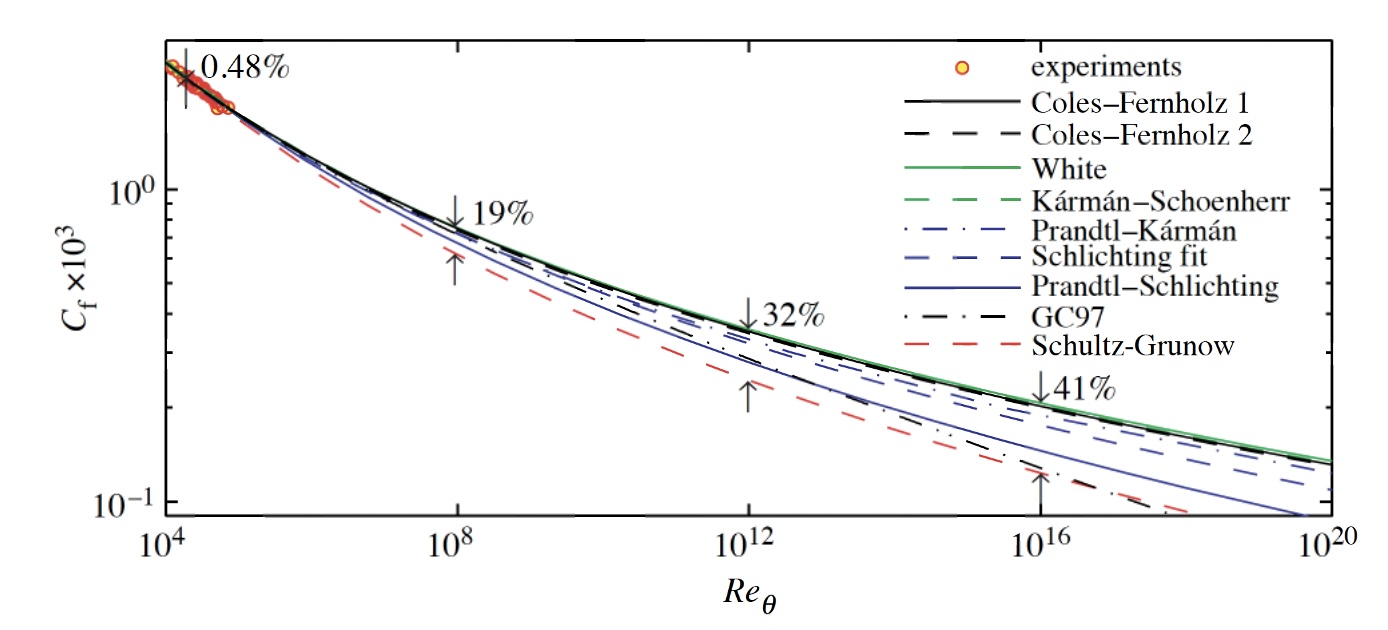}
        \caption{Reproduction of figure 2 of Nagib, Chauhan \& Monkewitz (2007) \cite{NCM07}:Asymptotic behaviour of modified $C_f$ relations and their local deviations.}
       \label{fig:fig3}
\end{figure}

Extrapolating all the ``modified'' relations up to $Re_{\theta}$ of $10^{20}$ in figure \ref{fig:fig3}, we find the differences to be 20\% at $Re_{\theta} = 10^8$ and about 40\% at $Re_{\theta} = 10^{16}$.  Even at ``galactic'' Reynolds numbers of $10^{16}$, the differences between all relations are comparable to their differences in the ``original form'' at $Re_{\theta}$ around $10^4$. At present, we believe that this relation best describes the skin friction coefficient towards asymptotically high Reynolds numbers. 

\begin{figure}
      \centering
      \includegraphics[width=0.75\textwidth]{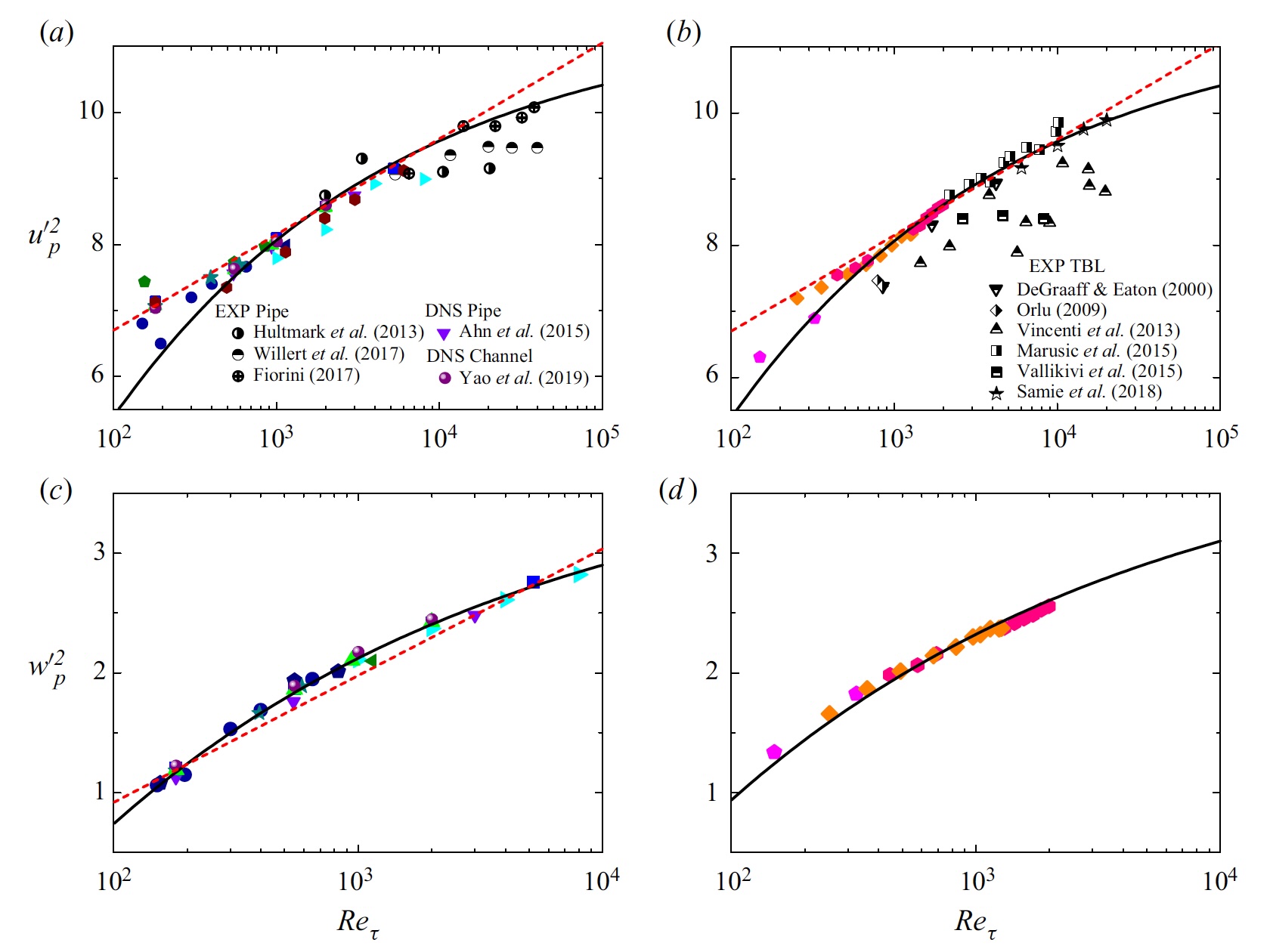}
        \caption{Reproduction of figure 2 of Chen \& Sreenivasan (2022) \cite{CS22}; note that the key to symbols in parts (c) and (d) are based on parts (a) and (b): The $Re_{\tau}$-variations of peak turbulence intensities. Streamwise intensity $u^{\prime 2}_p$ in channel and pipe (a), and in TBL (b). Spanwise intensity $w^{\prime 2}_p$ in channel and pipe (c), and in TBL (d). Newly included data are: EXP pipes of Princeton by Hultmark et al. (2012) \cite{hul12}, of CICLoPE by Willert et al. (2017) \cite{Willert} based on PIV measurement, and by Fiorini (2017) \cite{Fiorini} with hot-wire data corrected; DNS data of pipes by Ahn et al. (2015) \cite{Ahn} and of channels by Yao, Chen \& Hussain (2019) \cite{Yao19}; EXP data of TBL by DeGraaff \& Eaton (2000) \cite{Degraff}, \"Orl\"u (2009) \cite{Orlu}, Vincenti et al. (2013) \cite{Vincen}, Marusic et al. (2015) \cite{MCKH15}, Vallikivi et al. (2015) \cite{Valli} and Samie et al. (2018) \cite{sam18}; see figure legends for the corresponding symbols. Solid lines are fit to equation (2.2)\cite{CS22}, whose parameters are summarized in table~1\cite{CS22} Dashed lines indicate $u^{\prime 2}_p = 0.63~\ln(Re_{\tau} ) + 3.8$ by Marusic et al. (2017) \cite{MBH17}, and $w^{\prime 2}_p = 0.46~\ln(Re_{\tau})-1.2$ adopted by us for reference, both of which arise from the Gaussian-logarithmic model for high-order moments, as discussed in the text and shown in figures 4 and 5 of Chen \& Sreenivasan \cite{CS22}.}
        \label{fig:fig4}
\end{figure}

Peak in the streamwise fluctuation intensity: Recently, with the growth of the direct numerical simulations (DNS) data base, the high Reynolds number trends of streamwise fluctuations in wall-bounded turbulence have received increasing attention. One focus has been the ${Re}_\tau$-trend near the wall of the peak in intensity of the streamwise component of the fluctuation velocity, $<uu>^+_P$.  In the following, we compare the observed trends with the finite limit at infinite Reynolds number based on the boundedness of dissipation (Chen \& Sreenivasan \cite{CS22,CS23}) and the logarithmic unbounded growth as predicted by the attached eddy model founded on Townsend's hypothesis (e.g., Marusic \& Monty (2019) \cite{MM19}). %Two previous comparison regarding the different prediction of this trend were presented by Monkewitz, Nagib \& Boulanger (2917) \cite{MNB17} and by Nagib, Monkewitz \& Sreenivasan (2022) \cite{NMS22}. 
In particular, Chen \& Sreenivasan concluded that the near-wall peaks in the variance of the fluctuations are bounded for large $Re_\tau$ according to the relation 
\begin{equation}
 <uu>^+_P=11.5-19.3~Re_\tau^{-0.25}.   
\end{equation} 
The relevant results from their work are presented in figure \ref{fig:fig4}, with the figure caption providing the data sources.
In the spirit of the discussion on the skin friction, two coefficients in the proposed relation can be adjusted slightly to match those from Monkewitz \cite{M22}: 
\begin{equation}\label{eq:001}
    <uu>^+_P~=~11.3-17.7~Re_\tau^{-0.25}.
\end{equation}
%Marusic et al. (2017) \cite{MBH17} and Smits et al. (2021) \cite{Smits21} present experimental data and correlations with $Re_{\tau}$ based on the logarithmic trend of the attached eddy model. The most recent works are those of Monkewitz (2023) \cite{M22,M23} and Chen \& Sreenivasan (2023) \cite{CS22,CS23}.

%Figure 2

%Figure 3

%The relevant results from his work are presented in figure \ref{fig:fig5}, with the figure caption providing a complete explanation of data and fits.  Previously Monkewitz and Nagib \cite{MN15}, had estimated the peak streamwise stress for ZPG boundary layers as 
%\begin{equation}\label{eq:002}
%<uu>^+_P~=~22~-~\frac{340}%{U^+_\infty}, 
%\end{equation}
%but subsequently recognized that the data they used may have exaggerated the low $1/U^+_\infty$ (or high $Re_\tau$) limit of $22$.

%Using $U^+_\infty = (1/\kappa)\ln{R_\tau} + B$, the relation \ref{eq:002} is converted to a function of $Re_\tau$:
%\begin{equation}\label{eq:003}
    %<uu>^+_P~=~22-%\frac{340~\kappa}%{\ln{Re_\tau}} + H. O. T.
%\end{equation}

A formula from Monkewitz \& Nagib (2015) has the form 
\begin{equation}\label{eq:004}
    <uu>^+_P~=~12.8-\frac{80~\kappa}{\ln{Re_\tau}}.
\end{equation}

 The constants in (3) are somewhat different in the authors' original work. Their readjustment in the formula is done here only to explore if (3) can be brought in agreement with (2), over the limited Reynolds number range where reliable data are available, merely by readjusting the constants without changing the functional form.

%Figure 4

%Figure 5
\begin{figure}
      \centering
      \includegraphics[width=0.65\textwidth]{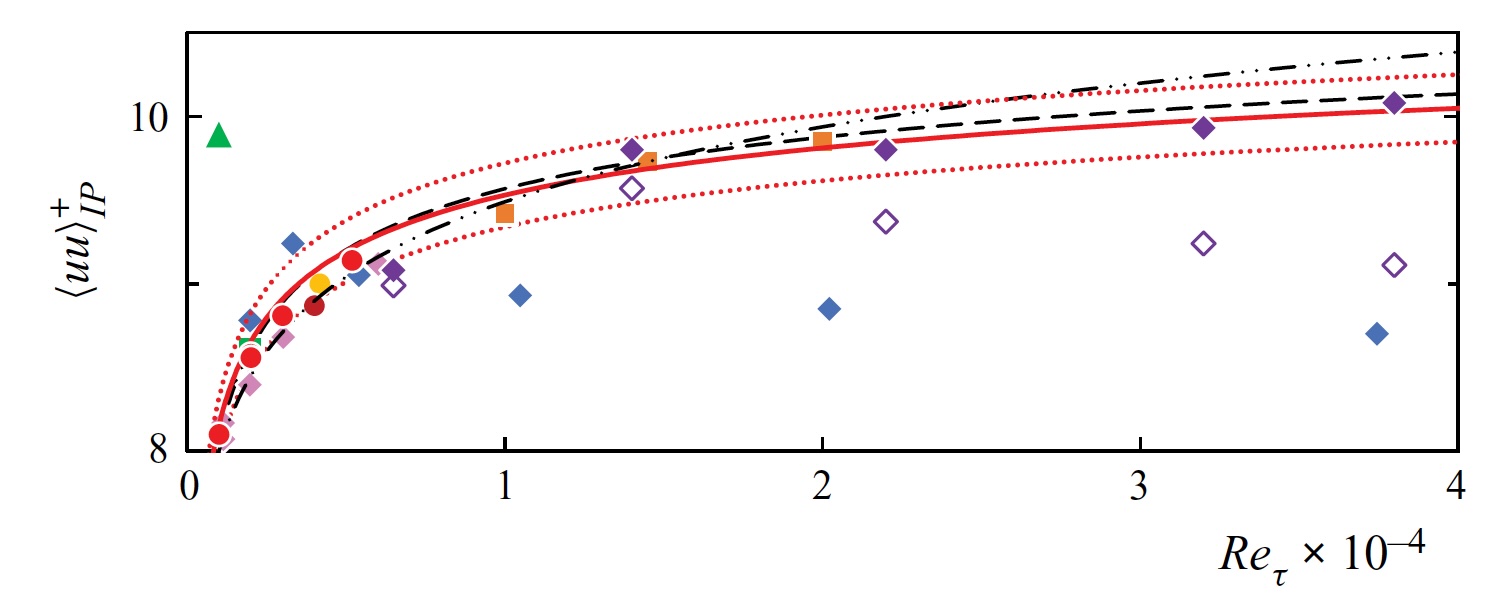}
        \caption{Reproduction of figure 2 of Monkewitz (2022) \cite{M22}: Inner peak height $<uu>^+_{IP}$ vs. $Re_{\tau}$: (red) ---, $\cdot \cdot \cdot$, (equation 2.4 in \cite{M22}) $ \pm 2 \%$; (black) - - -, $11.5 - 19.3~Re_{\tau}^{-0.25}$ of \cite{CS22}; (black) $- \cdot \cdot -$, $3.54 + 0.646~\ln Re_{\tau}$ of Samie et al. (2018) \cite{sam18} Channel DNS (circle): (red) DNS of table 1, (dark red) DNS of Bernardini, Pirozzoli \& Orlandi (2014) \cite{ber14}, (yellow) DNS of Lozano-Dur\'an \& Jim\'enez (2014)\cite{loz14} Pipe (diamond): (pink) DNS of Pirozzoli et al. (2021) \cite{pirozzoli}, (blue) Superpipe NSTAP data of Hultmark et al. (2012) \cite{hul12}, (purple) corrected and uncorrected (open diamond) CICLoPE hot-wire data of Fiorini (2017) \cite{Fiorini} ZPG TBL (square): (green) Sillero, Jim\'enez \& Moser (2013) \cite{sillero}, (orange) Samie et al. (2018) \cite{sam18} Couette (triangle): (green) Kraheberger et al. (2018) \cite{kra18a}.}
        \label{fig:fig5}
        \end{figure}

\begin{figure}
      \centering
      \includegraphics[width=0.7\textwidth]{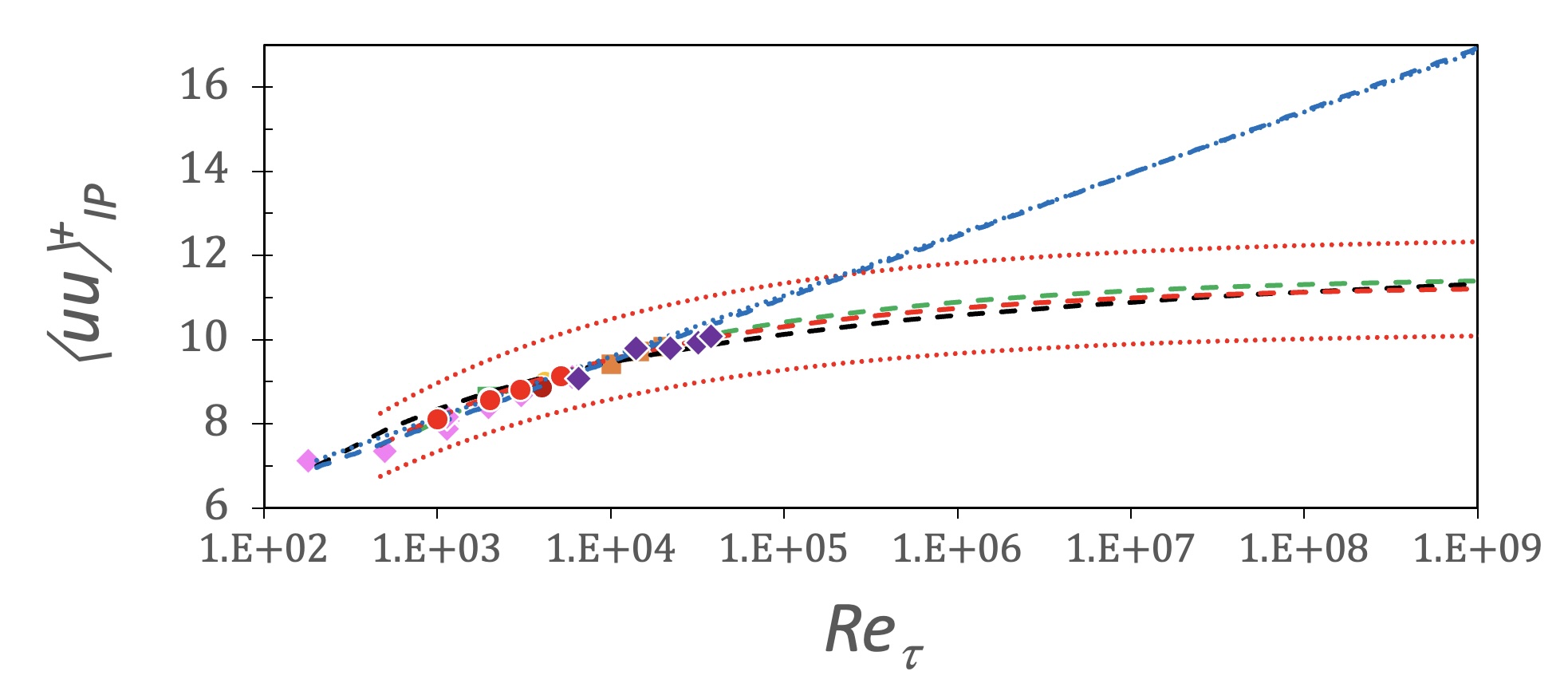}
        \caption{Comparison between trends of peak streamwise normal stress over wide range of Reynolds numbers (data from figure \ref{fig:fig5} excluding Couette \cite{kra18a}, NSTAP \cite{hul12} and uncorrected Fiorini \cite{Fiorini} results), and projections of equation \ref{eq:004} (black - - -) based on Monkowitz \& Nagib (2015) \cite{MN15}, relation  $<uu>^+_P = 11.5 - 19.3~Re_{\tau}^{-0.25}$ of Chen \& Sreenivasan (2022) \cite{CS22} (green - - -), and equation \ref{eq:001} of Monkewitz (2022) \cite{M22} (red - - - \& $\pm 10 \%$ $\cdot \cdot \cdot$), demonstrating that three proposed trends are in agreement. The proposed relations of Marusic et al. (2015) \cite{MCKH15}, equation \ref{eq:006} (blue - - -), and Samie et al. (2018)  \cite{sam18}, equation \ref{eq:007} (blue $\cdot \cdot \cdot$), deviate from agreement with other relations for $Re_\tau > 50,000.$}
        \label{fig:fig6}
\end{figure}

In figure \ref{fig:fig6}, equations \ref{eq:001} and \ref{eq:004} are compared over a wide range of $Re_\tau$ to the proposed relations of Marusic et al. (2015) \cite{MCKH15}:
\begin{equation}\label{eq:005}
    <uu>^+_P~=~0.63~\ln{Re_\tau} + 3.8,
\end{equation}
and the relation of Samie et al. (2018) \cite{sam18}:
\begin{equation}\label{eq:006}
    <uu>^+_P~=~0.646~\ln{Re_\tau} + 3.54.
\end{equation}

%Figure 6

\noindent It appears that the saturation formula fits the data somewhat better. There are other tests (see Nagib, Vinuesa \& Hoyas (2024) \cite{NVH24}, especially figures 3 and 5) that favor the saturation with the $-1/4$ power, but a direct test of the variance is only as good as in figure 4 here. This is the reason for the present effort to determine how high an $Re_\tau$ is required to see the differences directly and unambiguously. It should also be pointed that, while three plausible theoretical underpinnings of the saturation result were advanced in \cite{CS22, CS23}, each of them has an empirical content. An example of a promising direction of future research is the investigation of Smits et al. (2021) \cite{Smits21}, where the Reynolds number dependence of the leading coefficient of the Taylor expansion of $\langle uu\rangle^+$ about the wall is investigated among other quantities. The $Re_\tau$ dependence of the four Taylor coefficients for the channel in their Table 1 are least-squares fitted by
\begin{equation}\label{eq:008}
\frac{\langle uu\rangle^+}{(y^+)^2}\,(y^+\to 0) = 0.0157\,\ln{Re_\tau} + 0.0660
\end{equation}
with $R^2 = 0.991$ and an average deviation of 0.67\% (0.98\% maximum) from the fit.
The alternate fit, not explored by Smits et al. (2021) \cite{Smits21} is
\begin{equation}\label{eq:009}
\frac{\langle uu\rangle^+}{(y^+)^2}\,(y^+\to 0) = 0.246 - 0.400\,Re_\tau^{-1/4}
\end{equation}
with $R^2 = 1.000$ and an average deviation of 0.13\% (0.26\% maximum) from the fit. There is little doubt that the data support the asymptotic result of about 0.25 for $\frac{\langle uu\rangle^+}{(y^+)^2}$. 

\vspace{2cm}

Conclusions: Remarkable agreement is found in figure \ref{fig:fig6} over range $10^3 < Re_\tau < 10^9$ among the trends of the three proposed relations: Monkewitz (2022) \cite{M22} 
(equation \ref{eq:001}), the relation $<uu>^+_P = 11.5 - 19.3~Re_{\tau}^{-0.25}$ of 
Chen \& Sreenivasan (2022) \cite{CS22}, and the relation of Monkewitz \& Nagib (2015) \cite{MN15} with modified coefficients. The exception is the trend based on attached eddy model reported by Marusic et al. (2017) \cite{MCKH15} and by Samie et al. (2018) \cite{sam18} in equation \ref{eq:006}, both of which start to depart from other relations near $Re_\tau = 50,000$. While $Re_\tau$ of neutral atmospheric boundary layers do exceed $10^5$, the uncertainty in the measurements does not render them useful to resolve the question of the asymptotic limit. Thus we conclude that only wall-bounded turbulent flows with $Re_\tau$  $>$ O($10^6$), using accurate measurements of $<uu>^+_P$, can clearly and directly distinguish between recently proposed trends of the peak in streamsise normal stress profiles, i. e.,  $<uu>^+_{IP}$.    

Accurate measurements of high-order moments \cite{CS22,CS23}, with guidance from consistent asymptotic expansions, see, e.g., Monkewitz (2023) \cite{M23}, will be helpful in resolving the issue on hand.  Such diagnostics have been developed recently by Chen \& Sreenivasan (2024) \cite{CS24}, who have made detailed comparisons between the saturation formulae for high-order moments and experimental and DNS data. These comparisons show that the saturation forms are superior to the logarithmic formulae.  Establishing which of the two approaches is more valid is not only significant for estimating the streamwise normal stresses in the high Reynolds numbers range, but also for the fundamental understanding of many aspects of wall-bounded turbulence. 

What appears clear is also that the logarithmic and power scalings discussed earlier are mutually exclusive. This is seen by considering the transformation between competing large Reynolds number expansions of any quantity $F_i$ in terms of $\ln{Re_\tau}$ and of $Re_\tau^{-\gamma}$:
\begin{equation}\label{eq:007}
\left\{
\begin{array}{ll}
F_i = f_1^{\mathrm{log}} \ln{\mathrm{Re}}_{\tau, i} + f_2^{\mathrm{log}} & i=1,2 \\
F_i = f_1^{\mathrm{pow}} + f_2^{\mathrm{pow}} \mathrm{Re}_{\tau, i}^{-\gamma} & i=1,2
\end{array} \right\},
\quad ~~~
%\Rightarrow
f_1^{\mathrm{log}}(y^+) = f_2^{\mathrm{pow}}(y^+)\,\frac{\mathrm{Re}_{\tau, 1}^{-\gamma} - \mathrm{Re}_{\tau, 2}^{-\gamma}}{\ln{\mathrm{Re}_{\tau, 1}} - \ln{\mathrm{Re}_{\tau, 2}}}.
\end{equation}
It follows that the coefficient $f_1^{\mathrm{log}}(y^+)$ of $\ln{Re_\tau}$ in the logarithmic expansion goes to zero for any positive $\gamma$, as the two $\mathrm{Re}_{\tau, i}$ in equation \ref{eq:007} become large.

%\textbf{Data availability:} The data used for this paper can be obtained by contacting S. Hoyas at serhocal@mot.upv.es. 

%\begin{acknowledgments}

%\end{acknowledgments}
\newpage

\section*{References}\label{ref}
\bibliography{turbulence}% Produces the bibliography via BibTeX.

%\bibliographystyle{jfm}
%\bibliography{jfm-main}

\end{document}